\documentstyle[12pt,epsfig]{article}
\begin{document}
\thispagestyle{empty}
\begin{flushright}

\end{flushright}
\vskip 15pt

\begin{center}
{\Large {\bf Phenomenology of spinless adjoints in two Universal Extra
Dimensions}}
\renewcommand{\thefootnote}{\alph{footnote}}

\hspace*{\fill}

\hspace*{\fill}
 
{ \tt{
Kirtiman Ghosh\footnote{E-mail address:
kirtiman.ghosh@saha.ac.in}, Anindya Datta\footnote{E-mail address:
adphys@caluniv.ac.in}
}}\\

\small {\em Department of Physics, University of Calcutta,\\
92, A. P. C. Road, Kolkata 700009, India }
\\

\vskip 40pt

{\bf ABSTRACT}

\vskip 0.5cm

\end{center}
We discuss the phenomenology of $(1,1)$-mode adjoint
scalars in the framework of two Universal Extra Dimensions.  The
Kaluza-Klein (KK) towers of these adjoint scalars arise in the
4-dimensional effective theory from the 6th component of the gauge
fields after compactification. Adjoint scalars can have
KK-number conserving as well as KK-number violating interactions. We
calculate the KK-number violating operators involving these scalars
and two Standard Model fields. Decay widths of these scalars into
different channels have been estimated.  We
have also briefly discussed pair-production and single production of
such scalars at the Large Hadron Collider.

\vskip 30pt

\section{Introduction}
The primary aim for the next generation particle physics experiments
will be to find out whether a new dynamics beyond the Standard Model
(SM) really exists around the TeV scale of energy. A great effort have
been put in also to pin down the exact nature of this new dynamics at
the TeV Scale.  In this endeavour, lots of attention have been paid to
the theories with one or more extra space like dimensions. These extra
dimensional theories can be classified into two main classes. In one
of these, the standard model fields are confined to a (3 + 1)
dimensional subspace of the full manifold. Models of ADD \cite{add} or
RS \cite{rs} fall in this category.  On the other hand, there are
class of models where some or all of the SM fields can access
the full space-time manifold.  One such example is Universal Extra
Dimension (UED), where all the fields can propagate in the full
manifold \cite{UED}. Apart from the rich phenomenology, UED models in
general offer possible unification of the gauge couplings at a
relatively low scale of energy, not far beyond the reach of the next
generation colliders \cite{unificUED}.  Moreover, particle spectra of
UED models contain a weakly interacting stable massive particle, which
can be a good candidate for cold dark matter \cite{dark_ued5,dark_ued6}.

Phenomenology of one UED (1UED, space time is $4 + 1$ dimensional), 
have been extensively studied in the context of low energy 
experiments \cite{lowUED} as well as
high energy collider experiments \cite{colliderUED}. In this article,
we will study some aspects of a particular variant of the UED model
where all the SM fields can access 5 space like and 1 time
like dimensions. This is called {\em two Universal Extra Dimension} (2UED)
Model which has few additional attractive features. 2UED model can
naturally explain the long life time for the proton decay \cite{dobrescu0}
and more interestingly predicts that the number of fermion generations
should be an integral multiple of three \cite{dobrescu1}.

Recently, signals of 2UED model in future colliders like LHC
\cite{dobrescu4, dobrescu3} and ILC \cite{kong} have been studied in some 
details. In this article, we will concentrate on the phenomenology of 
some of the scalars in this theory and their possible production at
the LHC.

In $(1 + 5)$ dimensional (6D) space time, gauge fields have 6
components. However, after compactification, $(1+3)$ dimensional 
(4D) effective theory comprises of usual SM gauge fileds along 
with their Kaluza-Klein (KK) excitations. The 6th component of 
the gauge fields emerge
as KK towers of scalar fields transforming as the adjoints of the
respective gauge groups. Each KK-mode fields in 2UED model is
specified by a pair of positive integers (called the {\em
KK-numbers}). Phenomenology of the $(1,1)$-mode adjoint scalars will
be discussed in this article.

The plan of the article is the following. We will give a brief
description of the model in the next section. Interactions of the 
adjoint scalars will be discussed in section 3. Section
4 will be devoted to the decays of the $(1,1)$-mode adjoint 
scalars. We will briefly discuss the possible production 
mechanism of these scalars in section 5. We summarise in 
the last section.

\section{Two Universal Extra Dimensions}

As the name suggests, in 2UED all the SM fields can propagate
universally in the six-dimensional space-time. Four space time
coordinates $x^{\mu}$ ($\mu=0,1,2,3$) form the usual
Minkowski space. Two transverse spacial dimensions of
coordinates $x^4$ and $x^5$ are flat and are compactified with 
$0\le~x^4,~x^5~\le L$. This implies that the extra dimensional
 space (before compactification) is a square with sides $L$. 
Identifying the opposite sides of the
square will make the compactified manifold a torus. However, toroidal
compactification, leads to 4D fermions that are vector-like with
respect to any gauge symmetry. The alternative is to identify 
two pairs of adjacent sides of the square:

\begin{equation}
 (y,0)~\equiv~(0,y),~~~(y,L)~\equiv~(L,y),~~~\forall~y\in~[0,L]
\label{b_cond}
\end{equation}
This is equivalent to folding the square along a
diagonal and gluing the boundaries. Above compactification mechanism
automatically leaves at most a single 4D fermion of definite chirality
as the zero mode of any chiral 6D fermion \cite{dobrescu2}.

The field values should be equal at the identified points, modulo
possible other symmetries of the theory. The physics at identified points is
identical if the Lagrangian takes the same value for any field
configuration:
\begin{eqnarray}
{\cal L}\vert_{x^\mu,y,0}\;=\;{\cal L}\vert_{x^\mu,0,y};~~
{\cal L}\vert_{x^\mu,y,L}\,=\,{\cal L}\vert_{x^\mu,L,y} \nonumber
\end{eqnarray}
This requirement fixes the boundary conditions for 6D scalar fields
 $\Phi(x^\alpha)$ and 6D Weyl fermions $\Psi_{\pm}(x^\alpha)$. The
 requirement that the boundary conditions for 6D scalar or fermionic
 fields are compatible with the gauge symmetry, also fixes the boundary
 conditions for 6D gauge fields. The folding boundary conditions do
 not depend on continuous parameters, rather there are only eight
 self-consistent choices out of which one particular choice leads to
 zero mode fermions after compactification. Any 6D field (fermion/gauge
 or scalar) $\Phi(x^\mu,x^4,x^5)$ can be decomposed as:

\begin{equation}
 \Phi(x^\mu,x^4,x^5)~=~\frac{1}{L}\sum_{j,k}f^{(j,k)}_n(x^4,x^5)
\Phi^{(j,k)}(x^\mu)
\label{kk_dcomp}
\end{equation}
Where,
\begin{equation}
 f^{(j,k)}_n(x^4,x^5)~=~\frac{1}{1~+~\delta_{j,0}\delta_{k,0}}\left[e^{-in\pi/2}cos\left(\frac{jx^4+kx^5}{R}~+~\frac{n\pi}{2}\right)~+~cos\left(\frac{kx^4-jx^5}{R}~+~\frac{n\pi}{2}\right)\right]
\label{kk_func}
\end{equation}
The compactification radius $R$ is related to the size, $L$, of the 
compactified space as : $L = \pi R$. Where 4D fields
$\Phi^{(j,k)}(x^\mu)$ are the $(j,k)$-th KK 
modes of the 6D field $\Phi(x^{\alpha})$ and $n$ is a integer
whose value is restricted to $0,1,2$ or $3$ by the boundary
conditions. It is obvious from the form of $f^{(j,k)}_n(x^4,x^5)$ that
only $n=0$ allows zero mode ($j=k=0$) fields in the 4D effective
theory. The zero mode fields and the interactions among zero modes can
be identified with the SM.

 The requirements of anomaly cancellation and fermion mass generation
 force the weak-doublet fermions to have opposite {\em 6D chiralities}
 with respect to the weak-singlet fermions. So the quarks of one
 generation are given by $Q_+~\equiv~(U_+,D_+),~U_-,~D_-$.
 The 6D doublet quarks and leptons decompose into a
 tower of heavy vector-like 4D fermion doublets with left-handed zero
 mode doublets.  Similarly each 6D singlet quark and lepton decompose
 into the towers of heavy 4D vector-like singlet fermions along with
 zero mode right-handed singlets. These zero mode fields are
 identified with the SM fermions. As for example, SM doublet and
 singlets of 1st generation quarks are given by
 $(u_{L},d_{L})~\equiv~Q_{+L}^{(0,0)}(x^{\mu})$, $u_{R}~\equiv~U_{-R}^{(0,0)}(x^{\mu})$
 and $d_{R}~\equiv~D_{-R}^{(0,0)}(x^{\mu})$.

 In 6D, each of the gauge fields, has six
 components. Upon compactification, they decompose into  towers of 4D
 spin-1 fields, a tower of spin-0 fields which are eaten by heavy
 spin-1 fields. Another tower of 4D spin-0 fields, all belonging to
 the adjoint representation of the corresponding gauge group, remain
 in the physical spectrum. These are the physical {\em spinless
 adjoints} in which we are interested.

\begin{figure}[h]
\begin{center}
\epsfig{file=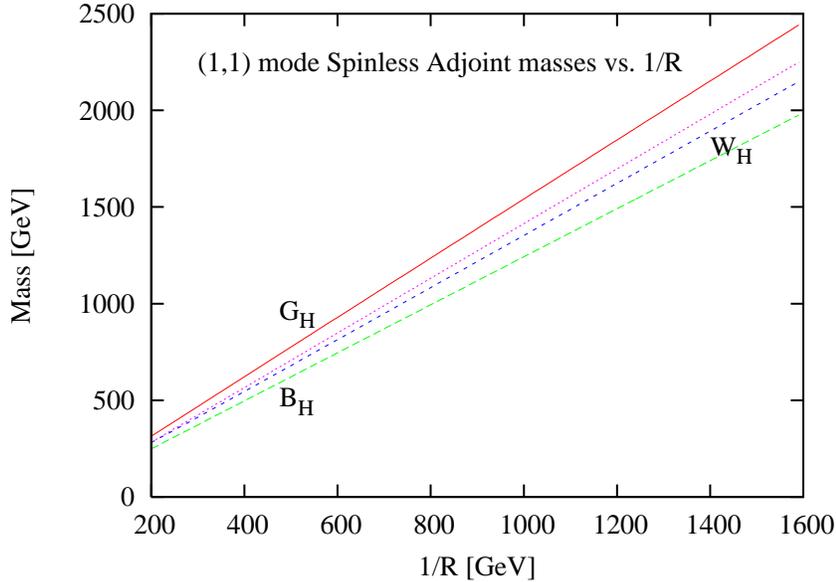,width=8cm,angle=270}
\end{center}
\caption{Variation of $G^{(1,1)}_H ({\rm solid}), W^{3(1,1)}_H$ (small dash) and  $B^{(1,1)}_H$ (large dash)
masses (after incorporating the one loop radiative corrections) with
$R^{-1}$.  For comparison, $\frac{\sqrt{2}}{R}$ (dotted) has also
been presented.}
\label{mass}
\end{figure}

The tree-level masses for $(j,k)$-th KK-mode particles are given by
$\sqrt{M_{j,k}^2~+~m_{0}^{2}}$, where $M_{j,k}=\sqrt{j^2+k^2}/R$.
 $m_0$ is the mass of the corresponding zero mode particle. As a result,
 the tree-level masses are approximately degenerate. This degeneracy
 is lifted by radiative effects. 
 The fermions receive mass corrections from the gauge interactions 
 (with gauge bosons and adjoint scalars) and Yukawa interactions. 
All of these give positive mass shift. The gauge fields and spinless 
adjoints receive mass corrections from the self-interactions and gauge
 interactions. Gauge interactions with fermions give a negative mass 
shift. While the self-interactions give positive mass shift with 
different strength with respect to the former. However, masses of the 
hypercharge gauge boson $B_{\mu}^{(j,k)}$ and the corresponding 
scalar $B_{H}^{(j,k)}$ 
receive only negative corrections from fermionic loops. Numerical 
computation shows that the lightest KK particle is the spinless adjoint 
$B_{H}^{(1,0)}$, associated with the hypercharge gauge boson. As a result, 
2UED model gives rise to a scalar dark matter candidate. As an 
illustrative example, we have plotted (in Fig.\ref{mass}) the variation 
of $(1,1)$-mode adjoint scalar masses (after including the radiative 
corrections) with $R^{-1}$. For comparison, we have also plotted 
$\frac{\sqrt{2}}{R}$, which is the tree-level mass of the $(1,1)$-th 
KK-mode particles.

Conservation of momentum (along the extra dimensions) in the full
theory, implies KK number conservation in the effective 4D
theory. Beginning with the SM-like interactions in the 6D,
(called the {\em bulk interactions}) one can obtain the {\em
KK-number} as well as {\em KK-parity conserving} interactions, in
4D effective theory after compactification.  However,
one can generate KK number violating operators
at one loop level, starting from the bulk interactions. Structure of 
the theory demands that these 
operators can only be on $(0,0,),~(0,L)$ and $(L,L)$ points of the 
{\em chiral square}. 
Bulk interactions being symmetric under KK-parity, operators generated by
loops also conserve the KK-parity. These KK number violating
operators are phenomenologically very important. A single non zero
mode KK particle can be produced only via the KK number violating 
interactions. Thus the complete 4D effective Lagrangian can be written as:
\begin{eqnarray}
 {\cal L}_{4D}&=& \int_{0}^{L}dx^{4}\int_{0}^{L}dx^{5}\{{\cal{L}}_{bulk} + \delta(x_{4})\delta(L - x_{5}){\cal{L}}_{2} + [ \delta(x_{4})\delta(x_{5}) + \nonumber \\ & + &
\delta(L - x_{4})\delta(L - x_{5})]{\cal{L}}_{1}\}
\label{6action}
\end{eqnarray}
 ${\cal{L}}_{bulk}$ includes 6D 
kinetic term for the quarks, leptons, $SU(3)_{C}\times SU(2)_{L}\times U(1)_{Y}$ 
gauge fields, a Higgs doublet, 6D Yukawa interactions of the quarks 
and leptons to the Higgs doublet, and a 6D Higgs potential. 
${\cal{L}}_1$ and ${\cal{L}}_2$ contain KK number violating interactions. 
As for example, the lowest dimensional localized operator that involve 
$-ve$ chirality 6D quark field ($F_{-}$) appear in ${\cal{L}}_{p}$ 
(p = 1,2) as
\begin{equation}
\frac{C_{pF}}{2^{2}M_{s}^{2}}i\bar F_{-R}\Gamma^{\mu}D_{\mu}F_{-R} +
(\frac{C_{pF}^{\prime}}{2^{2}M_{s}^{2}}i\bar F_{-R}\Gamma
^{l}D_{l}F_{-L} + h.c )
\label{fixed_op}
\end{equation}
where $\Gamma^{\mu}$ with $\mu = 0,1,2,3$ and $\Gamma ^{l}$ with $l =
4,5$ are six anti-commuting $8\times8$ matrices, $D^{\mu} , D^{l}$ are
6D covarient derivative, $C_{pF} , C_{pF}^{\prime}$ are dimensionless
parameters, and $M_{s}$ is the cut-off scale. 
${\cal{L}}_{1}$ and ${\cal{L}}_2$ also include 4D like kinetic terms
for all 6D scalar and gauge fields and some part of 6D kinetic
term. Contributions to those localized operators might be induced by
physics above the cut-off scale. We assume that those UV generated
localized operators are also symmetric under KK parity, so that the
stability of the lightest KK particle which can be a promising dark
matter candidate, is ensured. Loop contributions by the physics below
cut-off scale $M_s$ are used to renormalize the localized
operators. These contributions are calculated in \cite{1_loop} at one
loop order. Assuming bare contributions at the cutoff scale $M_s$ can
be neglected, RG evolution fixes the values of the $C$ parameters.

\section{Interactions of adjoint scalars}

In this section, we will discuss the possible interaction of a
$(j,k)$-mode adjoint scalar. The interactions of spinless adjoints can
be classified as KK number conserving and KK number violating
interactions. KK number conserving interactions arise from the
compactification of the bulk Lagrangian, where as, KK number violating
interactions arise mainly due to the loops involving KK number
conserving interactions.

\subsection{KK-number conserving interactions}

  Tree-level interactions of adjoint scalars with zero and non-zero
  mode fermions arise from the 6D kinetic terms for the fermions.
After compactification and integrating over the compactified co-ordinates
one can obtain the 4D effective interactions: 

\begin{eqnarray}
{\cal L}~\supset~&-&ig \delta_{0,1,3}^{(j_1,k_1)(j_2,k_2)(j_3,k_3)} r_{j_2,k_2}^{*} r_{j_3,k_3} \bar\psi_{+L}^{(j_1,k_1)} A_{H}^{(j_2,k_2)} \psi_{+R}^{(j_3,k_3)} \nonumber\\
&-& ig \delta_{0,1,3}^{(j_1,k_1)(j_2,k_2)(j_3,k_3)} r_{j_2,k_2}^{*} r_{j_3,k_3} \bar\psi_{-R}^{(j_1,k_1)} A_{H}^{(j_2,k_2)} \psi_{-L}^{(j_3,k_3)} \nonumber\\
&+& h.c                           \label{H_int_F}
\end{eqnarray}
$g$ is the gauge coupling for the gauge group in consideration and 
$r_{j,k}=(j+ik)/\sqrt{j^{2}+k^{2}}$. The above
form of the interactions are valid both for $A_{H}^{(j,k)}$ being an
abelian spinless adjoint field and non-abelian spinless adjoint with
the replacement of $A_{H}^{(j,k)}~\rightarrow~T^{a}A_{H}^{a(j,k)}$.
$T^{a}$'s are the gauge group generators corresponding to the
representation of $\Psi_{\pm}$. Appearance of the the
$\delta$-functions\footnote{We follow the same convention as in the
ref.\cite{dobrescu2}.} in the above expression ensure the conservation
of KK-number. We list the relevant Feynman Rules arising from the
above interactions in Fig.\ref{rules_kkconsv} of Appendix A.

\subsection{KK-number violating interactions}

Starting with the KK number conserving bulk interactions of 2UED one
can generate KK number violating operators via loop effects. As for
example, a dimension 5 operator involving two zero mode fermions and a even
KK parity (with $j+k$ even) $(j,k)$- mode spinless adjoint can be constructed
in such a way. 

We have listed the KK-number violating 2-point and 3-point functions
in Fig.\ref{rules_KKviolat} of Appendix A. The amplitude for $A_{H}^{(j,k)}\to
\psi_{-R}^{(0,0)}\psi_{-R}^{(0,0)}$ is also calculated in Appendix B and
is given by,
\begin{equation}
{\cal M}=-i\frac{g}{M_{j,k}}\;(K_{C_{1 \psi_{R}} C_{2 \psi_{R}}}^{(j,k)}
- K_{C_{1\psi_{R}}^{\prime} C_{2\psi_{R}}^{\prime}}^{(j,k)}) \;
\left[ \bar u(p_1)p\!\!/ P_{R}u(p_2)\right]
\label{ffA_amp}
\end{equation}
Defining $ \tilde C^{f}_{j,k}~=~K_{C_{1\psi_{R}}
C_{2\psi_{R}}}^{(j,k)}-K_{C_{1\psi_{R}}^{\prime}
C_{2\psi_{R}}^{\prime}}^{(j,k)}$, one can parametrise $\tilde
C^{f}_{j,k}=\tilde \xi_{f} \frac{1}{16\pi^2}
Log(M_{s}^{2}/M_{j,k}^{2})$. $\tilde \xi_{f_{L,R}}$ are
dimensionless parameters which fix the couplings of adjoint scalars
with two zero mode fermions.

\begin{eqnarray}
\tilde \xi_{q_L}&=& 4g_{s}^{2}+\frac{9}{4}g^{2}+\frac{3}{4}g^{\prime 2}Y_{q_L}^{2} \nonumber   \\
\tilde \xi_{q_R}&=& 4g_{s}^{2}+\frac{3}{4}g^{\prime 2}Y_{q_R}^{2} \nonumber   \\
\tilde \xi_{e_L}&=& \frac{9}{4}g^{2}+\frac{3}{4}g^{\prime 2}Y_{e_L}^{2} \nonumber   \\
\tilde \xi_{e_R}&=& \frac{3}{4}g^{\prime 2}Y_{e_R}^{2}
\label{xi} 
\end{eqnarray}
where $Y_{f}$ is the hypercharge of the corresponding fermion $f$.
Spinless adjoints can also couple with two SM gauge bosons. These
couplings are generated from finite 1-loop diagrams (Appendix B). The
dimension-5 operators, involving a $(1,1)$ mode colour spinless
adjoint and two SM gauge boson, are given by:

\begin{equation}
\textit{C}_{\gamma g}^{G}\;\epsilon^{\mu\nu\alpha\beta}\;
Tr\left[G_{\mu\nu}\;A_{\alpha\beta}\;G^{(1,1)}_{H}\right] +
\textit{C}_{Zg}^{G}\epsilon^{\mu\nu\alpha\beta}Tr[G_{\mu\nu}Z_{\alpha\beta}G^{(1,1)}_{H}]
\label{su3}
\end{equation}
The dimension-5 operators, involving a $(1,1)$ mode $U(1)$ or $SU(2)$
spinless adjoint $(V^{(1,1)}_H)$ ($B_{H}^{(1,1)}~or~W_{H}^{3(1,1)}$) 
and two SM gauge bosons, are given by,

\begin{eqnarray}
&~&\textit{C}_{gg}^{V}\epsilon^{\mu\nu\alpha\beta}\;
Tr\left[G_{\mu\nu}G_{\alpha\beta}V^{(1,1)}_{H}\right] + 
\textit{C}_{\gamma \gamma}^{V}\epsilon^{\mu\nu\alpha\beta}
A_{\mu\nu}A_{\alpha\beta}V^{(1,1)}_{H} \nonumber \\
&+&\textit{C}_{\gamma Z}^{V}
\epsilon^{\mu\nu\alpha\beta}A_{\mu\nu}Z_{\alpha\beta}V^{(1,1)}_{H} + 
\textit{C}_{ZZ}^{V}\epsilon^{\mu\nu\alpha\beta}Z_{\mu\nu}Z_{\alpha\beta}
V^{(1,1)}_{H} \nonumber \\
&+&\textit{C}_{W^{+}W^{-}}^{V}\epsilon^{\mu\nu\alpha\beta}W^{+}_{\mu\nu}W^{-}_{\alpha\beta}V^{(1,1)}_{H}
\label{su2}
\end{eqnarray}

where $G_{\mu \nu},~A_{\mu \nu},~Z_{\mu \nu}$ and $W^{\pm}_{\mu \nu}$
are the field strengths of gluon, photon, $Z$-boson and
$W^{\pm}$-boson respectively. The values of the $C$ coefficients can
be found in Appendix B. We have used $\alpha = 1/128$ and
$sin^{2}\theta_{W}=0.23$. The SM running of strong
coupling constant (with $\alpha_s = 0.1$ at a scale of 1 TeV) has 
been used. Yukawa couplings for all light quarks ($u,~d,~c,~s$) 
have been neglected. Top and bottom quark Yukawa couplings are 
taken to be 1 and 0.02 respectively. For quarks the resulting 
values of $\tilde \xi$ parameters at a scale about $0.5 ~\rm TeV$ are $\tilde \xi_{q_L}=6.82$,
$\tilde \xi_{u_R}=6.02$ and $\tilde \xi_{d_R}=5.89$.

Following the same algorithm in Appendix-B, one can obtain 
dimension-6 operators involving two spinless adjoints and a 
gauge boson of the form:
\begin{eqnarray}
&~&A_{\mu \nu}\partial^{\mu}B_{H}^{(1,1)}\partial^{\nu}W_{H}^{3(1,1)} + 
Tr\left[G_{\mu \nu}\partial^{\mu}B_{H}^{(1,1)}\partial^{\nu}G_{H}^{(1,1)}\right] \nonumber\\
&+& Tr\left[G_{\mu \nu}\partial^{\mu}G_{H}^{(1,1)}\partial^{\nu}W_{H}^{3(1,1)}\right]
+Z_{\mu \nu}\partial^{\mu}B_{H}^{(1,1)}\partial^{\nu}W_{H}^{3(1,1)}
\end{eqnarray}
However, first, second and the third operator vanish identically 
if we use equation of motion for photon and gluon respectively. 
Moreover the decay of $W_{H}^{3(1,1)}$ to a $B_{H}^{(1,1)}$ (can take
place via the last one) and a massive SM gauge boson is kinematically 
forbidden for $R^{-1}$ values upto 1.5 TeV.

\section{Decays of $(1,1)$-mode spinless adjoints}
Until now the discussions about the couplings of the adoint scalars
were more or less general about any $(j,k)$ mode adjoint. However,
from this section we will focus specifically on the phenomenology of
the $(1,1)$ mode adjoints.

 $(1,1)$-mode spinless adjoint can decay into a lighter
$(1,1)$-mode particle (kinematically possible only for $SU(3)$ and
$SU(2)$ spinless adjoints) and one or two SM particles via the
KK-number conserving couplings. This can also decay to a pair of SM
fermions/gauge bosons via the KK-number violating interactions. In
this section we compute the all such decay branching fractions of $(1,1)$-
mode spinless adjoints using the interactions derived in the previous section.

\subsection{Decays of $G_{H}^{(1,1)}$}

As discussed, $G_{H}^{(1,1)}$ can decay to the SM quarks via loop
induced operators. The amplitude for $G_{H}^{(1,1)}~\to~\bar q q$ is
given by

\begin{equation}
{\cal {M}}~=~-i\frac{g_s}{M_{j,k}} \bar u(p_1)p\!\!/ \left(\tilde
C^{q_R}_{j,k}P_{R} + C^{q_L}_{j,k}P_{L}\right) u(p_2)
\label{qqgh}
\end{equation}

Applying Dirac equation it is easy to see that the amplitude is
proportional to quark mass. So $G_{H}^{(1,1)}$ predominantly decays
into $t \bar t$ for $M_{G_{H}^{(1,1)}} > 2 m_t $. However, there are
other dimension-5 operators which couple $G_{H}^{(1,1)}$ with a SM
gluon and an electro weak gauge bosons. Since these couplings arise
from finite 1-loop diagrams (Fig.\ref{amp_ggH}), they are 
suppressed by a logarithm
compared to the $G_{H}^{(1,1)}q \bar q$ couplings. $G_{H}^{(1,1)}$
being heavier than the $U(1)$ gauge boson $B_{\mu}^{(1,1)}$, can 
decay into $B_{\mu}^{(1,1)}$ and a SM gluon. This
coupling is also generated by finite 1-loop diagram and is thus suppressed.

We first consider the decay into $ t \bar t $ and $b \bar b$. The
widths can be computed in terms of the parameters $\tilde \xi_{q_L}$ and
$\tilde \xi_{q_R}$ given in Eq.\ref{xi}. The decay width into $ q \bar q $
is given by
\begin{eqnarray}
\Gamma(G_{H}^{(1,1)}\to q \bar q)&=&\Gamma_{0}^{G_{H}}(\tilde
\xi^{2}_{q_L}+\tilde
\xi^{2}_{q_R})(\frac{m_q}{m_{G_{H}^{(1,1)}}})^{2}(1-\frac{4m_{q}^{2}}{m^{2}_{G_{H}^{(1,1)}}})^{\frac{1}{2}}
\label{width_qqgh}
\end{eqnarray}
Where, $q$ can be $t$ or $b$ and $\Gamma_{0}^{G_{H}}~=~\frac{\alpha_{s}}{4}m_{G_{H}^{(1,1)}}\tilde C_{1,1}^{2}$ \\

The decay width to $g B_{\mu}^{(1,1)},~gZ$ and $g \gamma$, induced by
finite 1-loop effect, are as the following :
\begin{eqnarray}
\Gamma(G_{H}^{(1,1)}\to B_{\mu}^{(1,1)} g)&=&\frac{\alpha_{s}^{2}\alpha}{32 \pi^{2}cos^{2}\theta_{w}} (\sum_{\psi_{\pm}} \frac{Y_{\psi}}{2}\sigma_{\psi}A_{\psi}^{G B_{\nu}^{(1,1)} g})^{2} f_{G}(m_{B_{\nu}^{(1,1)}})  \nonumber \\
\Gamma(G_{H}^{(1,1)}\to gZ)&=&\frac{\alpha_{s}^{2}\alpha}{32 \pi^{2}cos^{2}\theta_{w}sin^{2}\theta_{w}} f_{G}(m_{z}) \nonumber \\
&\times& (\sum_{\psi_{\pm}} [I^{3}_{\psi}-Q_{\psi} sin^{2}\theta_{w}]\sigma_{\psi}A_{\psi}^{GgZ})^{2} \nonumber \\
\Gamma(G_{H}^{(1,1)}\to g \gamma)&=&\frac{\alpha_{s}^{2}\alpha}{32
  \pi^{2}} (\sum_{\psi_{\pm}} Q_{\psi} \sigma_{\psi}A_{\psi}^{G g
  \gamma})^{2} f_{G}(0)
\label{width_aagh}
\end{eqnarray}

Where $f_{G}(m)~=~[(m^{2}_{G_{H}^{(1,1)}}-m^{2})/m_{G_{H}^{(1,1)}}]^{3}$ and 
$A_{\psi}$'s (in Eqs.\ref{width_aagh}, \ref{aabh} and 
\ref{aaw3h}) are defined in Eqs.\ref{A_psi}. It
important to notice that for $M_{G_{H}^{(1,1)}} < 350 ~GeV $ the
dominant decay mode is $gZ$ rather than $b \bar b$. Although the
$G_{H}^{(1,1)}b \bar b$ couplings is logarithmically enhanced but it is
suppressed by $b$-quark mass.

\begin{figure}
\begin{center}
\epsfig{file=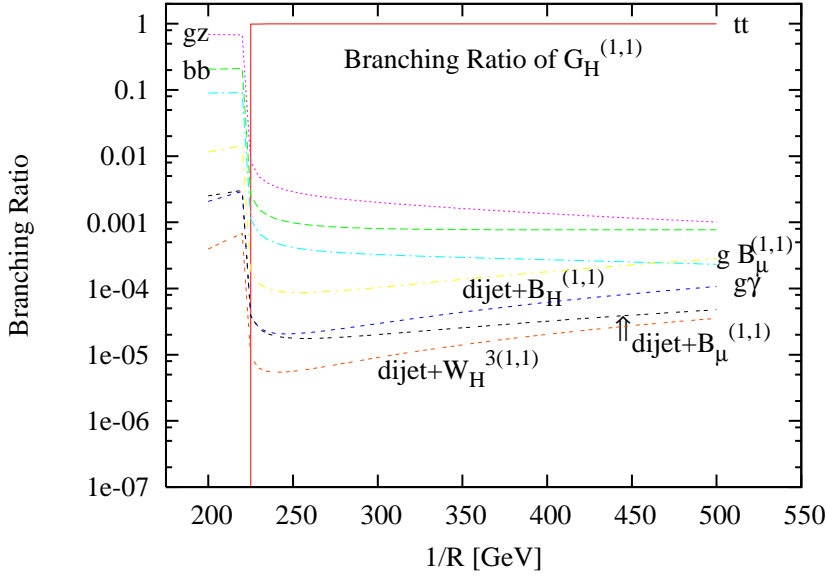,width=8cm,angle=270}
\end{center}
\caption{Branching Ratios of $G_H^{(1,1)}$. $M_s R = 10$
has been assumed in the calculation  }
\label{ghbranch}
\end{figure}

Beside those two body decays, $G_{H}^{(1,1)}$ undergoes tree-level
3-body decays to $B_{H}^{(1,1)},~B_{\mu}^{(1,1)}$ or $W_{H}^{3(1,1)}$
and SM fermion anti-fermion pairs. Branching fractions of those decays
are also presented in the Fig.\ref{ghbranch}

\subsection{Decays of $B_{H}^{(1,1)}$}

$B_{H}^{(1,1)}$ is the lightest $(1,1)$ mode KK particle. So, the
decay $B_{H}^{(1,1)}$ via KK number conserving interactions are not
kinematically allowed. It can decay to $f \bar f$ via dimension-5
operator in Eq.\ref{ffA_amp}. Since the coupling is proportional to the fermion mass,
$B_{H}^{(1,1)}$ decay predominantly to $t \bar t$ for
$m_{B_{H}^{(1,1)}}> 2 m_t$. The decay modes $b \bar b$ and $\tau \bar
\tau$ are suppressed due to fermion mass. $B_{H}^{(1,1)}$ can also decay
to two SM gauge bosons through dimension-5 operators Eq.\ref{su2}, generated
from finite 1-loop contribution. However, these vertices are suppressed by
a logarithm compared to $B_{H}^{(1,1)}f \bar f$ vertex.\\ The decay
width to $t \bar t,~b \bar b$ and $\tau \bar \tau$ are given by

\begin{eqnarray}
\Gamma(B_{H}^{(1,1)}\to t \bar t)&=&3\Gamma_{0}^{B_{H}}(\frac{1}{4}\tilde \xi^{2}_{t_L}+4\tilde \xi^{2}_{t_R})(\frac{m_t}{m_{B_{H}^{(1,1)}}})^{2}(1-\frac{4m_{t}^{2}}{m^{2}_{B_{H}^{(1,1)}}})^{\frac{1}{2}} \nonumber  \\
\Gamma(B_{H}^{(1,1)}\to b \bar b)&=&3\Gamma_{0}^{B_{H}}(\frac{1}{4}\tilde \xi^{2}_{b_L}+\tilde \xi^{2}_{b_R})(\frac{m_b}{m_{B_{H}^{(1,1)}}})^{2}(1-\frac{4m_{b}^{2}}{m^{2}_{B_{H}^{(1,1)}}})^{\frac{1}{2}} \nonumber \\
\Gamma(B_{H}^{(1,1)}\to \tau \bar
\tau)&=&\Gamma_{0}^{B_{H}}(\frac{9}{4}\tilde \xi^{2}_{\tau_L}+9\tilde
\xi^{2}_{\tau_R})(\frac{m_\tau}{m_{B_{H}^{(1,1)}}})^{2}(1-\frac{4m_{\tau}^{2}}{m^{2}_{B_{H}^{(1,1)}}})^{\frac{1}{2}}
\label{qqbh}
\end{eqnarray}
Where, $\Gamma_{0}^{B_{H}}=\frac{\alpha}{18 cos^{2}\theta_{w}}m_{B_{H}^{(1,1)}}\tilde C_{1,1}^{2}$. The decay widths of $B_{H}^{(1,1)}$ into two SM gauge bosons are as follows:
\begin{eqnarray}
\Gamma(B_{H}^{(1,1)}\to g g)&=&\frac{\alpha_{s}^{2}\alpha}{8 \pi^{2}cos^{2}\theta_{w}} (\sum_{\psi_{\pm}} \frac{Y_{\psi}}{2} \sigma_{\psi}A_{\psi}^{Bgg})^{2} f_{B}(0)  \nonumber \\
\Gamma(B_{H}^{(1,1)}\to \gamma \gamma)&=&\frac{\alpha^{3}}{16 \pi^{2}cos^{2}\theta_{w}} (\sum_{\psi_{\pm}} \frac{Y_{\psi}}{2} Q_{\psi}^{2}\sigma_{\psi}A_{\psi}^{B \gamma \gamma})^{2} f_{B}(0) \nonumber \\
\Gamma(B_{H}^{(1,1)}\to \gamma Z)&=&\frac{\alpha^{3}}{8 \pi^{2}cos^{4}\theta_{w}sin^{2}\theta_{w}} f_{B}(m_{Z}) \nonumber \\
 &(&\sum_{\psi_{\pm}} \frac{Y_{\psi}}{2}[I^{3}_{\psi}-Q_{\psi} sin^{2}\theta_{w}]Q_{\psi} \sigma_{\psi}A_{\psi}^{B \gamma Z})^{2} \nonumber \\
\Gamma(B_{H}^{(1,1)}\to ZZ)&=&\frac{\alpha^{3}}{16 \pi^{2}cos^{6}\theta_{w}sin^{4}\theta_{w}}f_{B}^{\prime}(m_{Z}) \nonumber \\ 
&(&\sum_{\psi_{\pm}} \frac{Y_{\psi}}{2}[I^{3}_{\psi}-Q_{\psi} sin^{2}\theta_{w}]^{2} \sigma_{\psi} A_{\psi})^{2}  \nonumber \\
\Gamma(B_{H}^{(1,1)}\to W^{+}W^{-})&=&\frac{\alpha^{3}}{32 \pi^{2}cos^{2}\theta_{w}sin^{4}\theta_{w}} f_{B}^{\prime}(m_{W})\nonumber \\
&(&\sum_{\psi_{\pm}} \frac{Y_{\psi}}{2}\sigma_{\psi} A_{\psi})^{2} 
\label{aabh}
\end{eqnarray}

\begin{figure}
\begin{center}
\epsfig{file=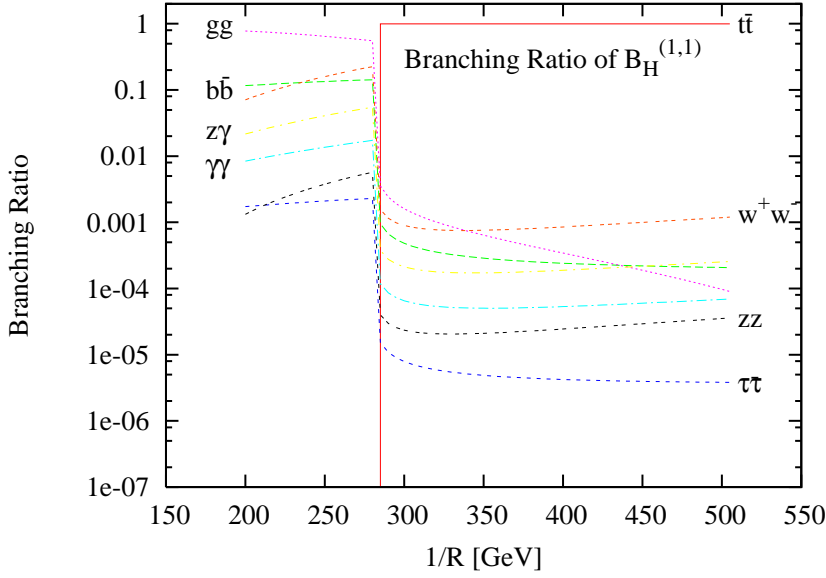,width=8cm,angle=270} 
\caption{Branching Ratios of $B_H^{(1,1)}$ to pair of SM particles. $M_{s} R = 10$
has been assumed in the calculation  }
\label{bhbranch}
\end{center}
\end{figure}

Where, $f_{B}^{\prime}(m)~=~m_{B_{H}^{(1,1)}}^{3}[1-4
m^{2}/m_{B_{H}^{(1,1)}}^{2}]^{3/2}$, and 
$f_{B}(m)=[(m^{2}_{B_{H}^{(1,1)}}-m^{2})/m_{B_{H}^{(1,1)}}]^{3}$. 
For $M_{B_{H}^{(1,1)}} < 2m_t $, $B_H$ dominantly decays to a pair of
gluons. For $M_{B_{H}^{(1,1)}} > 2 m_t $, beside $t \bar t$, the next
dominant decay mode is $W^{+}W^{-}$. This is a consequence of large
mass splitting of $(1,0)$ mode quarks and leptons as discussed in
Appendix B. The different branching ratios of $B_H$ are presented 
in Fig.\ref{bhbranch}.
\subsection{Decays of $W_{H}^{3(1,1)}$}

$W^{3(1,1)}_{H}$ is the next to the lightest $(1,1)$ mode particle. 
$W^{3(1,1)}_{H}$ can decay only into a pair of 
SM particles via KK number violating
interactions mentioned before. The dominant decay mode is 
again into $t \bar t$ for
$m_{W_{H}^{3(1,1)}}> 2 m_t$. $W^{3(1,1)}_{H}$ can decay to other SM
fermion anti-fermion pairs but such decays are suppressed
by the respective fermion masses. The decay width of $W^{3(1,1)}_{H}$
into $f \bar f$ is given by

\begin{equation}
\Gamma(W_{H}^{3(1,1)}\to f \bar f)~=~C_{F}\Gamma_{0}^{W^{3}_{H}}\tilde
\xi^{2}_{f_L}
(\frac{m_f}{m_{W_{H}^{3(1,1)}}})^{2}(1-\frac{4m_{f}^{2}}{m^{2}_{W_{H}^{3(1,1)}}})^{\frac{1}{2}}
\label{ffw3h}
\end{equation}
Where, $\Gamma_{0}^{W^{3}_{H}}=\frac{\alpha I_{3}^{2}}{2 Sin^{2}\theta_{w}}m_{W_{H}^{3(1,1)}}\tilde C_{1,1}^{2}$ and $C_{F}$ is the quadratic casimir. 

\begin{figure}
\begin{center}
\epsfig{file=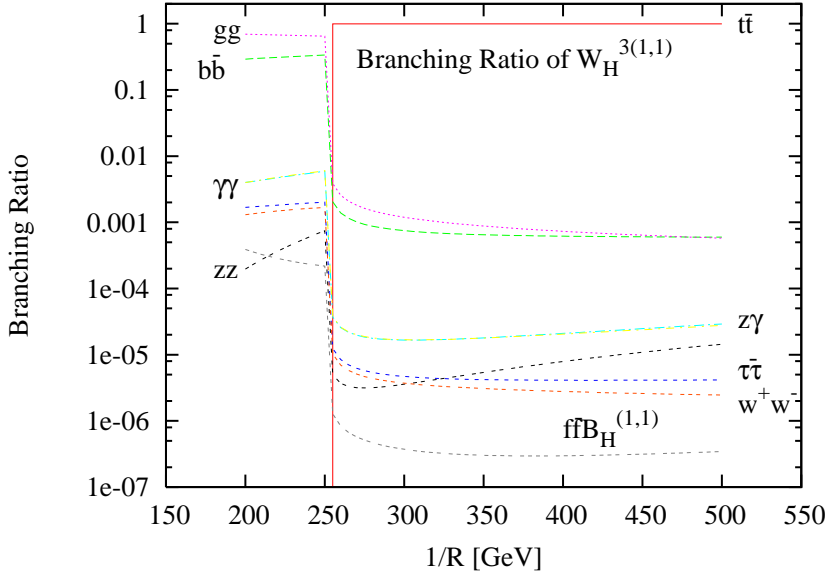,width=8cm,angle=270} 
\caption{Branching ratios of $W^{3(1,1)}_H$. $M_s R = 10$
has been assumed in the calculation  }
\label{branchwh}
\end{center}
\end{figure}

Apart from this decay, electrically neutral $(1,1)$ mode $SU(2)$ spinless adjoint can decay into two SM gauge bosons but these decays are again suppressed by a logarithm. The decay widths are given by
\begin{eqnarray}
\Gamma(W_{H}^{3(1,1)}\to g g)&=&\frac{\alpha_{s}^{2}\alpha}{8 \pi^{2}sin^{2}\theta_{w}} (\sum_{\psi_{+}} I_{\psi}^{3} \sigma_{\psi}A_{\psi}^{Wgg})^{2} f_{W}(0)  \nonumber \\
\Gamma(W_{H}^{3(1,1)}\to \gamma \gamma)&=&\frac{\alpha^{3}}{16 \pi^{2}sin^{2}\theta_{w}} (\sum_{\psi_{+}} I_{\psi}^{3} Q_{\psi}^{2}\sigma_{\psi}A_{\psi}^{W \gamma \gamma})^{2} f_{W}(0) \nonumber \\
\Gamma(W_{H}^{3(1,1)}\to \gamma Z)&=&\frac{\alpha^{3}}{8 \pi^{2}cos^{2}\theta_{w}sin^{4}\theta_{w}}  f_{W}(m_Z)\nonumber \\
 &(&\sum_{\psi_{+}} I_{\psi}^{3}[I^{3}_{\psi}-Q_{\psi} sin^{2}\theta_{w}]Q_{\psi} \sigma_{\psi}A_{\psi}^{W\gamma Z})^{2} \nonumber \\
\Gamma(W_{H}^{3(1,1)}\to ZZ)&=&\frac{\alpha^{3}}{16 \pi^{2}cos^{4}\theta_{w}sin^{6}\theta_{w}} f_{W}^{\prime}(m_Z) \nonumber \\ 
&(&\sum_{\psi_{+}} I_{\psi}^{3}[I^{3}_{\psi}-Q_{\psi} sin^{2}\theta_{w}]^{2} \sigma_{\psi} A_{\psi}^{WZZ})^{2} \nonumber \\
\Gamma(W_{H}^{3(1,1)}\to W^{+}W^{-})&=&\frac{\alpha^{3}}{32
  \pi^{2}sin^{6}\theta_{w}} (\sum_{\psi_{+}} I_{\psi}^{3}\sigma_{\psi}
A_{\psi}^{W W^{+}W^{-}})^{2} f_{W}^{\prime}(m_W) 
\label{aaw3h}
\end{eqnarray}
The previous definition of $f$ and $f^{\prime}$ functions holds here with
 proper change in spinless adjoint mass.
The branching ratios are presented in Fig.\ref{branchwh}.

There are also tree-level 3-body decay of $W_{H}^{3(1,1)}$ into
left-handed SM fermion anti-fermion pairs and $B_{H}^{(1,1)}$. As can
be seen from Fig.\ref{branchwh}, those decay modes are very suppressed
compared to others.

\section{Production of $(1,1)$-mode Spinless Adjoints}

We will first discuss the pair production of $(1,1)$-mode adjoint
scalars.  Production of $(1,0)$-mode scalars (in particular the
$G_{H}^{(1,0)}$) was discussed in ref.\cite{dobrescu2}. We will also
discuss the production of electroweak ($W^3 _H$ and $B_H$)
$(1,1)$-mode scalars along with $G_H$. Coupling of a pair of
$G_{H}^{(1,1)}$ with a zero mode gluon, and coupling of a (1,1) mode
quark with a zero mode quark and a $G_{H}^{(1,1)}$ arise from bulk
interaction. We have estimated the following cross-sections :
$\sigma(G_H G_H)$, $\sigma(G_H B_H)$, $\sigma(G_H W^3_H)$,
$\sigma(B_H B_H)$, $\sigma(W^3_H W^3_H)$, $\sigma(W^3_H B_H)$ in proton
proton collision at the LHC energies. CTEQ4L parton distribution functions
\cite{cteq4} are used to numerically evaluate the above
cross-sections. We have fixed the factorisation scale (for parton
distribution functions) and scale of $\alpha_s$ (where relevant) at 
${(1,1)}$ mode mass.

\begin{figure}
\begin{center}
\epsfig{file=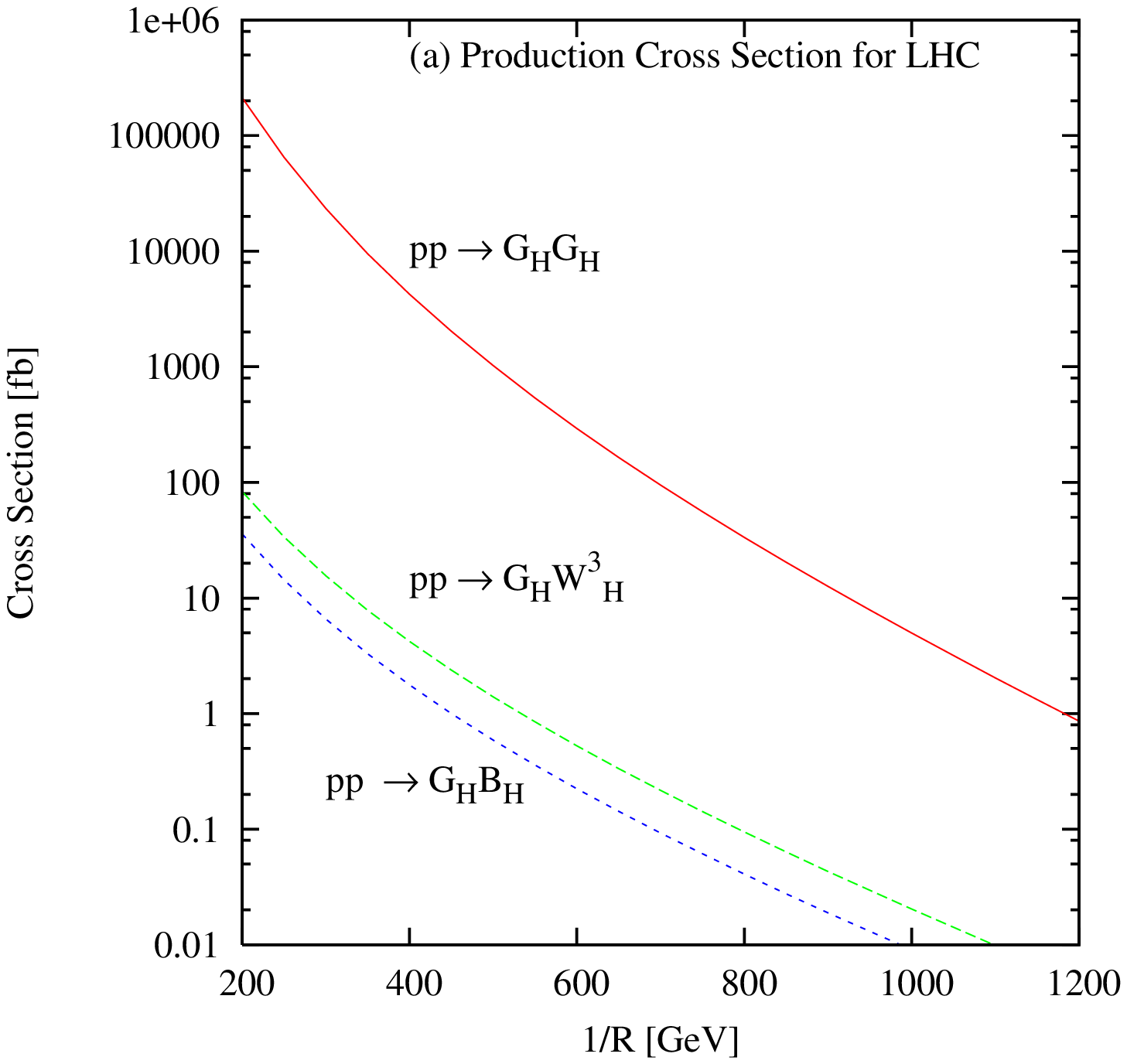,width=8cm,height=8cm,angle=270}
\epsfig{file=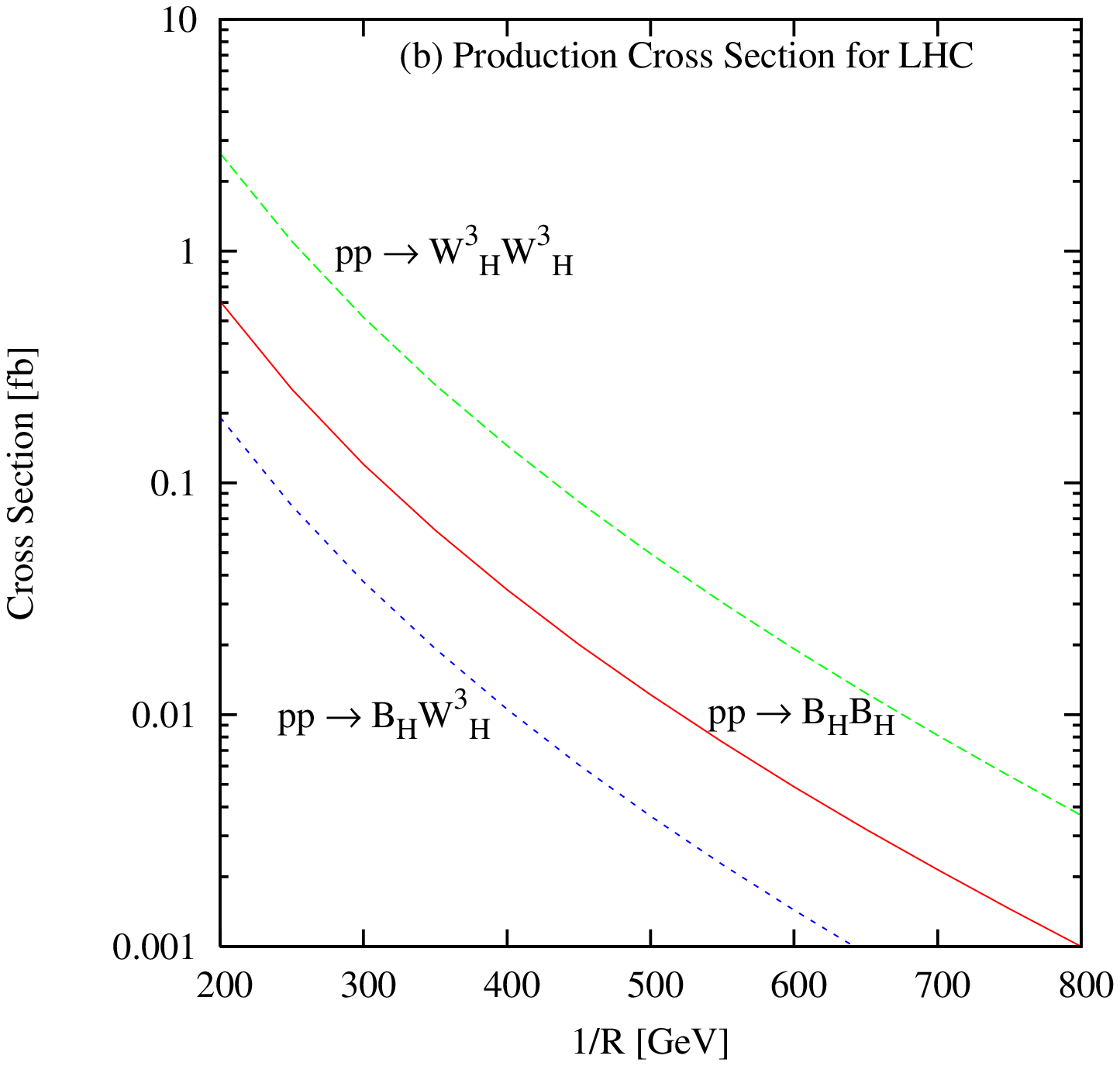,width=8cm,height=8cm,angle=270} 
\caption{Pair production cross-sections for various $(1,1)$ mode adjoint scalars. 
CTEQ4L parton distributions are used to evaluate the cross-sections.   }
\label{pair_prod}
\end{center}
\end{figure}

All the above cross-sections are presented in Fig.\ref{pair_prod}.
$G_{H}^{(1,1)}$ pair-production being a pure QCD process has a 
large cross-section at the LHC.  Single $B_{H}^{(1,1)}$ or
$W_{H}^{(1,1)}$ production along with a $G_{H}^{(1,1)}$ also have
large cross-sections.  On the other hand, pair productions of 
electroweak adjoint scalars are miniscule even for lower values of
$R^{-1}$. Dominance of $G_{H}^{(1,1)} G_{H}^{(1,1)}$ pair production
can be primarily attributed to contributions from the the gluon
gluon initiated contributions. Gluonic contributions are absent in 
all other cases we have presented in above two figures. Apart from the
$G_{H}^{(1,1)} G_{H}^{(1,1)}$ pair production, all other processes are
only initiated by quarks and an anti-quarks. LHC, being a proton
proton collider, anti-quarks can only arise from sea-excitation. Their
densities also fall sharply with adjoint scalar masses.

$G_{H}^{(1,1)}$ pair production varies from a few pb to few fb, as we
change $R^{-1}$ over a range from 200 to 1200 GeV.  Once produced
$G_{H}^{(1,1)}$ will decay dominantly to $t \bar t$, thus copiously
producing 4 $t$-quarks. Distinguishing this signal from the SM
background will be a challenging task. However, for $G_{H}^{(1,1)}$ masses
below $2 m_t$, it can decay to $g Z$, thus producing a spectacular
2-jet + 4 lepton signal.

Production cross-sections of $W_{H}^{(1,1)}$ and $B_{H}^{(1,1)}$ in
association with $G_{H}^{(1,1)}$ are also presented in
Fig.\ref{pair_prod}a. $W_{H}^{(1,1)}$ ($B_{H}^{(1,1)}$)
cross-sections varies from 100 (10) fb to 0.01 fb as we vary 
$R^{-1}$ from 200 to 1200 (1000) GeV.

Fig.\ref{pair_prod}b, shows the pair production of electroweak adjoint
scalars, namely $(1,1)$ mode of $W_H$ and $B_H$. These cross-sections
are small and are comparable with the single productions of these
scalars (discussed in the following).

Pair production of all these adjoint scalars via KK-number conserving
interactions, result in $4t$ signal, once the scalar masses are
greater than twice the top mass.

We will now briefly discuss the single production of adjoint
scalars, $W^{(1,1)}_{3H}$ and $B^{(1,1)}_{H}$, the adjoints of the
electro-weak gauge group. These can be produced at the LHC via gluon gluon
fusion as well as in association with top-quark.  However,
$G^{(1,1)}_{H}$ cannot be produced via gluon gluon fusion due to
$SU(3)$ symmetry.  The only single production mode for this strongly
interacting adjoint scalar is in association with a pair 
of $t$-quarks\footnote{Spin-less adjoints can be produced from production and
subsequent cascade decays of $V^{(1,1)}_\mu$.  The cross-sections for
these production channels
\cite{dobrescu3} are higher than the processes we are considering in
the following. However, in ref.\cite{dobrescu3}, spinless adjoints are
produced along with several number of jets. This makes their detection
at a hadronic collider difficult.}. This associated production rate is
proportional to the fifth power of gauge couplings. As a result, the
cross-sections, even for relatively lower values of $R^{-1}$, are
not so promising.  They also fall sharply with $R^{-1}$. This is
primarily due to the direct $R^{-1}$ dependence of $G^{(1,1)}_H t
\bar t$ couplings.

Resonance production cross-section of $W^{(1,1)}_{3H}$ and $B^{(1,1)}_{H}$ 
from $pp$ collision is given by 

\begin{equation}
\sigma (pp \rightarrow V^{(1,1)}_H + X) = \frac{\pi ^2}{36 s m_{V_H}}\;
\Gamma(V^{(1,1)}_H \rightarrow gg) \int_\tau ^1 \frac{dx}{x}\; g(x,m^{2}_{V}) 
g(\frac{\tau}{x},m_{V}^{2})
\label{r_cross}
\end{equation} 
s is the $pp$ center-of-mass energy square, $\tau$ is a dimensionless
parameter : $\frac{m_{V_H}^2}{s}$. $g$'s are the gluon densities inside a 
proton. 

\begin{figure}
\begin{center}
\epsfig{file=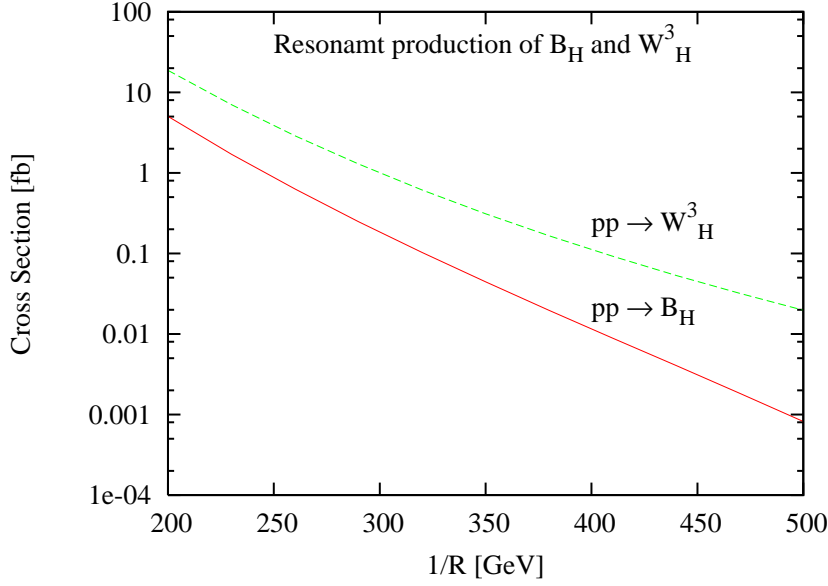,width=8cm,angle=270}
\end{center}
\caption{Resonant production cross-sections of $W^{3(1,1)}_{H}$ and 
$B^{(1,1)}_{H}$ at the LHC with $R^{-1}$.}
\label{prod_1}
\end{figure}

In the previous section, we obtained the expressions for the various
decay widths of the spinless adjoints. It is now straightforward to
calculate the cross-sections using the above expression. We have
presented the $W^{(1,1)}_{3H}$ and $B^{(1,1)}_{H}$ production
cross-sections in Fig.\ref{prod_1}.  CTEQ4L parton distributions
have been used to numerically evaluate the cross-sections.
Smallness of the cross-sections can be attributed to the narrow decay
widths of these scalars into a gluon pair. The narrow decay widths of
these electroweak adjoint scalars are evident in view of incomplete
anomaly cancellation as explained in the Appendix B.

\section{Conclusion}
To summarise, we have investigated the phenomenology of the adjoint
scalars in an effective 4D theory, resulting from the compactification
of Standard Model in 6 space time dimensions. These scalars arise from
the 6th component of the gauge fields of the gauge groups $SU(3)$,
$SU(2)$ and $U(1)$ respectively after compactification. Apart from
KK-number conserving interactions which arise from the bulk, adjoint
scalars have interactions with a pair of SM particles via KK-number 
violating but KK-parity conserving terms in the
interaction Lagrangian. The later couplings arise via one-loop effects
due to bulk interactions. Structure of the theory, in particular the
chiral nature of compactifiation forces these effective couplings to
be on the fixed points of the manifold. We have calculated these
effective couplings involving the $(j,k)$ mode of the adjoint scalars
($G_H$, $W_H$ and $B_H$) with a pair of SM fields. The possible decays
of $(1,1)$ mode scalars have been calculated. It is found that if kinematically
allowed, they will dominantly decay to a pair of the heaviest fermions,
namely the top-quark.  We have calculated the pair production
cross-section of the adjoint scalars in the context of Large Hadron
Collider.  Pair production of adjoint scalars involves only the
KK-number conserving interactions. $G^{(1,1)}_H G^{(1,1)}_H$,$G^{(1,1)}_H W^{3(1,1)}_H$, $G^{(1,1)}_H B^{(1,1)}_H$
cross-sections are large. On the other hand $B^{(1,1)}_H W^{3(1,1)}_H$,$W^{3(1,1)}_H W^{3(1,1)}_H$, $B^{(1,1)}_H
B^{(1,1)}_H$ pair productions at LHC are small. We have also computed the
single production rates of $W_{H}^{3(1,1)}$ and $B_{H}^{(1,1)}$ via
gluon gluon fusion which take place via
KK-number violating interactions, at the LHC.
 $G^{(1,1)}_H$ cannot be produced singly via gluon gluon
fusion. However, the single production
cross-sections via KK-number violating interaction are in general small.\\

{\bf Acknowledgements} KG acknowledges the support from Council of
Scientific and Industrial Research, Govt. of India. AD is partially
supported by Council of Scientific and Industrial Research, Govt. of
India, via a research grant 03(1085)/07/EMR-II.

\newpage
\renewcommand{\theequation}{A.{\arabic{equation}}}
\setcounter{equation}{0}

{\bf Appendix A : Relevant Feynman Rules} \\

\begin{figure}[h]
\epsfig{file=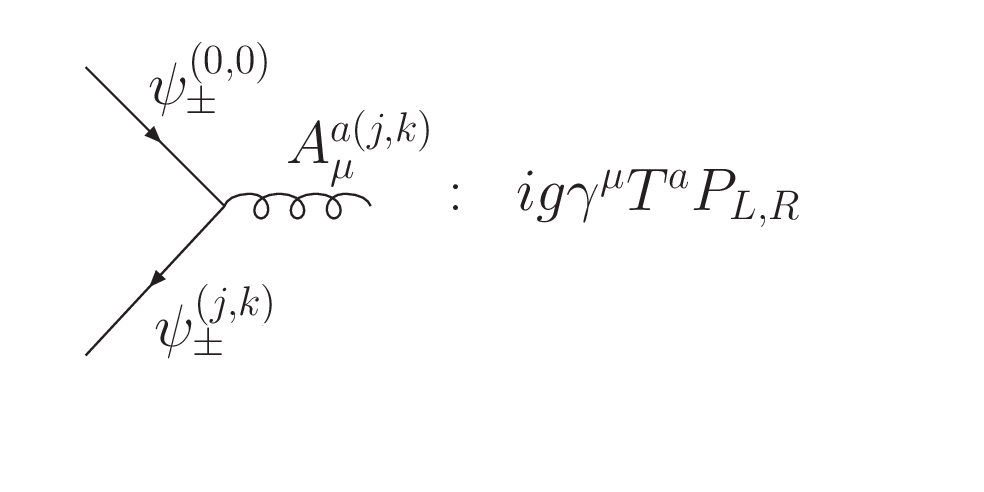,width=8cm}
\epsfig{file=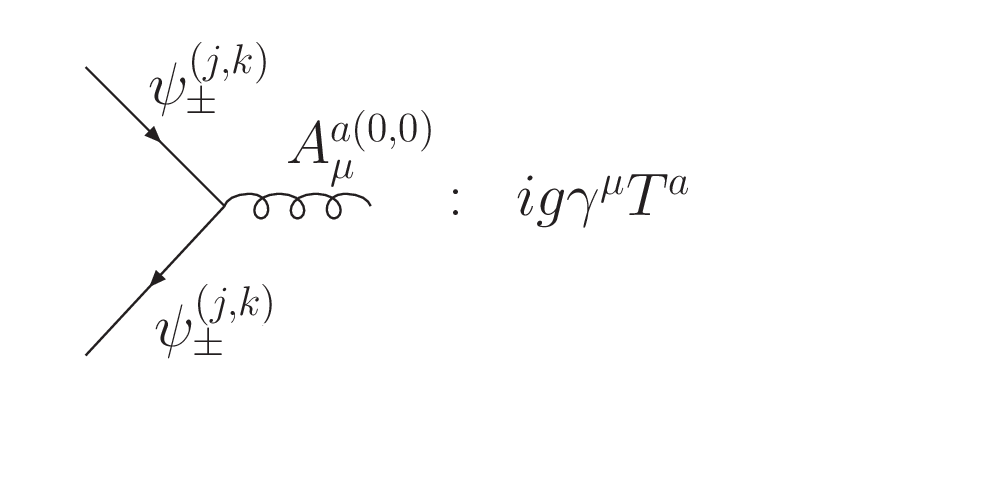,width=8cm}\\
\epsfig{file=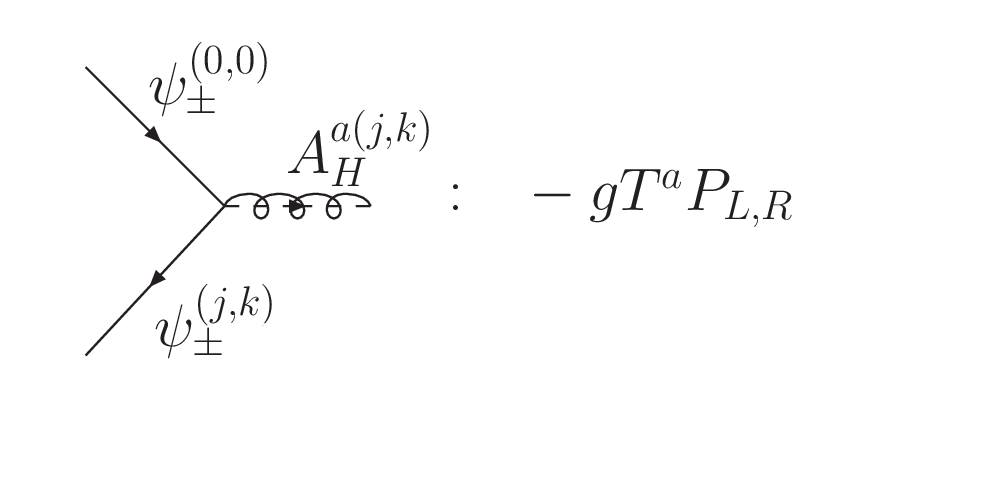,width=8cm}
\epsfig{file=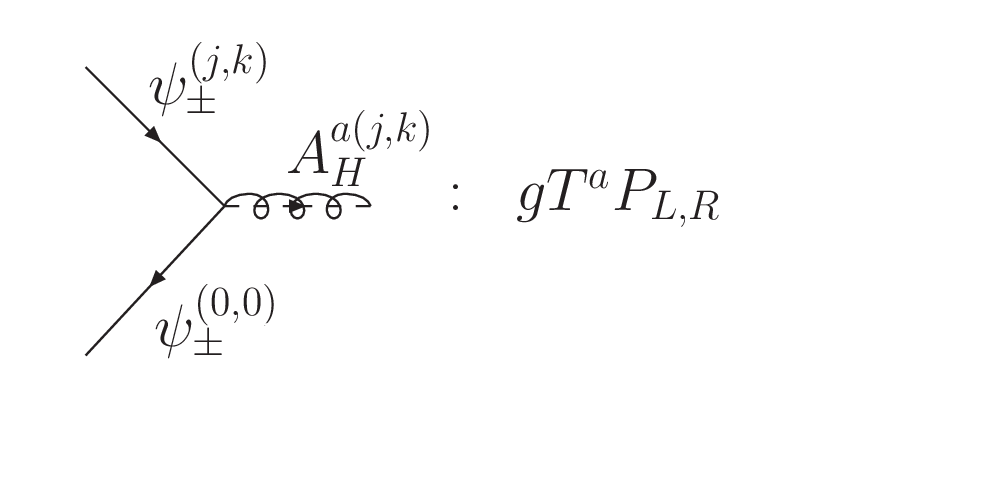,width=8cm}\\

\caption{Feynman rules of KK-number conserving interactions arising
from Eq.\ref{H_int_F}.  $g$ is the gauge coupling constant and $T^a$ is the
generator of the relevant gauge group. For $U(1)$, $T^a$s should be
replaced by $\frac{Y}{2}$. }
\label{rules_kkconsv}
\end{figure}

In this Appendix, we list the Feynman rules those are relevant for loop
calculations. KK number conserving vertices involving a gauge
boson or a spinless adjoint and two fermions are listed in Fig.\ref{rules_kkconsv}.

\begin{figure}[h]
\begin{center}
\epsfig{file=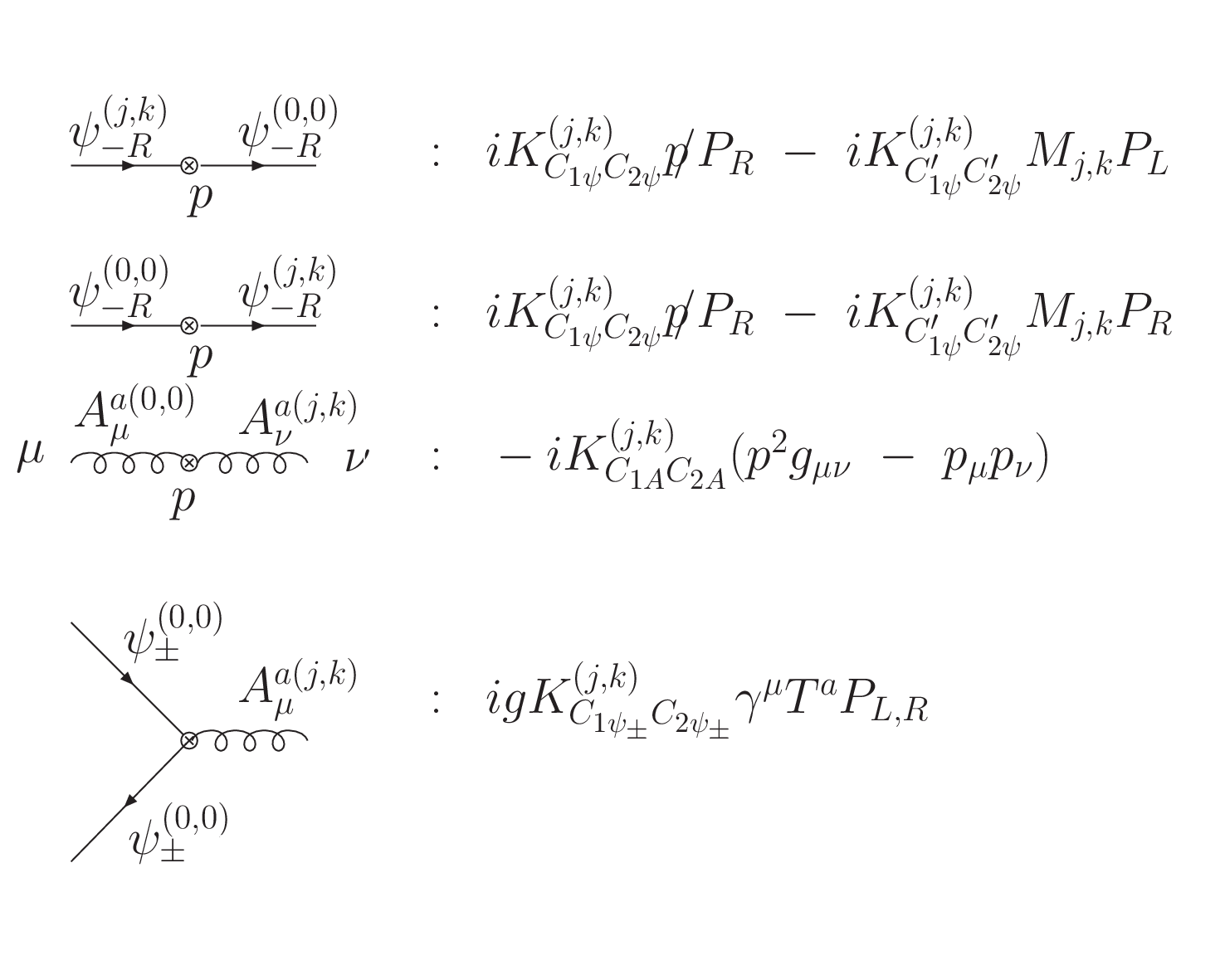,width=12cm}\\
\caption{Feynman rules of KK-number violating interactions arising from
 compactification of fixed point operators.}
\label{rules_KKviolat}
\end{center}
\end{figure}

Operators localized at the singular points, after compactification,
give rise to the KK number violating 2-point and 3-point
functions. These are listed in Fig.\ref{rules_KKviolat}. KK number
violating 2-point functions induce kinetic and mass mixing between
different $(j,k)$ modes. Corresponding 2-point and 3-point
functions involving electro-weak gauge bosons can be easily inferred
from those given in Fig.\ref{rules_KKviolat}.  $K^{(j,k)}_{C_1 C_2}$ 
in Fig.\ref{rules_KKviolat} is defined as:

\begin{equation}
K^{(j,k)}_{C_1 C_2}~=~\frac{2}{(\pi R M_s)^2 }(2 C_1~+~(-1)^{j}C_2)
\label{K}
\end{equation}

$C_1,~C_2$ are the dimensionless parameters, already introduced in
Eq.\ref{fixed_op}. In Appendix B, we will use these KK-number
violating 2-point functions to compute the coupling of an even
KK-parity spinless adjoint with two SM fermions.

\vspace*{.5in}

\renewcommand{\theequation}{B.{\arabic{equation}}}
\setcounter{equation}{0}

{\bf Appendix B : KK-number violating Loop Induced Couplings} \\

In this Appendix, we first compute $A_{H}^{(j,k)}\to
\psi_{-R}^{(0,0)}\psi_{-R}^{(0,0)}$ amplitude. This kind of
interactions are generated only by loop effects. One can construct dimension-5 operators
which couples two zero mode fermions and a $(j,k)$-mode (with $j+k$
even) spinless adjoint, using the Feynman rules in Fig.\ref{rules_kkconsv} and
Fig.\ref{rules_KKviolat}. As for example, the amplitude for $B_{H}^{(j,k)}\to \psi_{-R}^{(0,0)}\psi_{-R}^{(0,0)}$ is given by
\begin{equation}
{\cal
  M}=-ig^{\prime}\frac{Y_\psi}{2}\frac{1}{M_{j,k}}(K_{C_{1\psi_{R}}
  C_{2\psi_{R}}}^{(j,k)}-K_{C_{1\psi_{R}}^{\prime}
  C_{2\psi_{R}}^{\prime}}^{(j,k)}) \;\left[\bar u(p_1)p\!\!/ P_{R}u(p_2)\right]
\label{ffh_ver}
\end{equation}

\begin{figure}[h]
\begin{center}
\epsfig{file=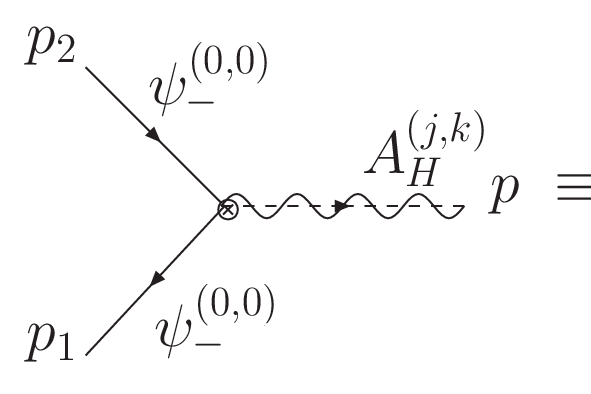,width=4cm}
\epsfig{file=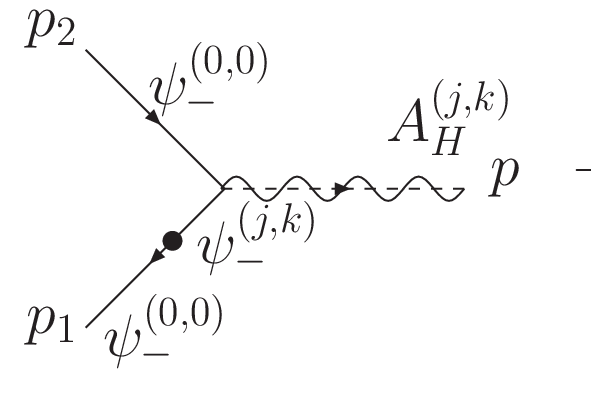,width=4cm}
\epsfig{file=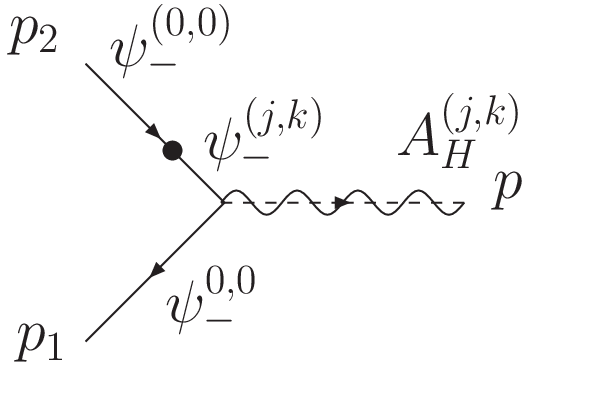,width=4cm}
 \caption{Vertices involving two zero mode fermions and a even KK parity
$(j,k)$ mode spinless adjoint}
\label{amp_ffH}
\end{center}
\end{figure}
\begin{figure}[h]
\begin{center}
\epsfig{file=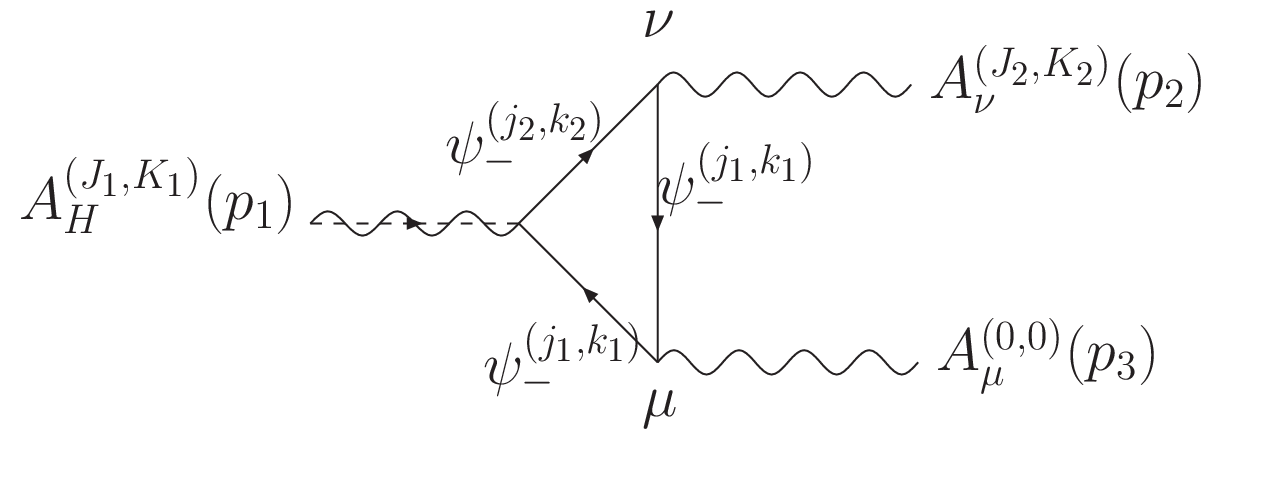,width=10cm}\\
\caption{Effective vertex involving one SM and an even KK parity
$(J_2,K_2)$ mode gauge boson with a even KK parity $(J_1,K_1)$ 
mode spinless adjoint.}
\label{amp_ggH}
\end{center}
\end{figure}

Modulo  the KK parity conservation, spinless adjoints can interact with two vector
modes via finite 1-loop diagram. The coupling of a $(J_1,K_1)$
mode spinless adjoint with a $(J_2,K_2)$ mode gauge boson and a zero mode
gauge boson is induced by the 1-loop diagram in Fig.\ref{amp_ggH}.\\

The amplitude for $A_{H}^{(J_1,K_1)}\to A_{\nu}^{(J_2,K_2)} 
A_{\mu}^{(0,0)}$ is given by:

\begin{equation}
{\cal M} =\frac{1}{4 \pi^{2}} \times (Gauge~Couplings)  \times
A_{\psi}\epsilon^{\mu \nu \alpha \beta}
\epsilon^*_{\nu}(p_2)\epsilon^*_{\mu}(p_3)p_{1\alpha}p_{2\beta}
\label{aah_ver}
\end{equation}
where we have defined $A_{\psi}$ in the following way
\begin{eqnarray}
A_{\psi}&=&\sum _{j_1,k_1} \sum _{j_2,k_2} A_{\psi}^{j_1,k_1;j_2,k_2}\nonumber \\
A_{\psi}^{j_1,k_1;j_2,k_2}&=& m_{\psi}^{j_1,k_1;j_2,k_2}(C_{12}-C_{11}-C_{0})- m_{\psi}^{j_2,k_2;j_1,k_1}(C_{12}-C_{11}) \nonumber \\
m_{\psi}^{j_1,k_1;j_2,k_2}&=&m_{j_1,k_1} Re[r^{*}_{J_1;K_1}r_{j_1,k_1}(\delta_{013}^{j_1,k_1;J_1,K_1;j_2,k_2}\delta_{103}^{j_2,k_2;J_2,K_2;j_3,k_3} \nonumber \\
&-&\delta_{013}^{j_2,k_2;J_1,K_1;j_1,k_1}\delta_{000}^{j_1,k_1;J_2,K_2;j_2,k_2})]
\label{A_psi}
\end{eqnarray}
$C_{0},~C_{11},~C_{12}$ are the scalar and vector three point
functions of `t Hooft, Passarino and Veltman \cite{veltman} and 
{\em ``Gauge Couplings''} in Eq.\ref{aah_ver}
corresponds to the product of three gauge couplings (arising in
Fig.\ref{amp_ggH}) times respective group theory factors. $C$
functions depend only on the three external masses and the three
internal (fermionic) masses. The coefficient
$m_{\psi}^{j_1,k_1;j_2,k_2}$ survives for a finite set of 
$(j_1,k_1;j_2,k_2)$. As for example, only $(1,0)$-mode
fermion contributes to the loop in the coupling of a
$(1,1)$ mode spinless adjoint with two zero-mode gauge bosons. 
 The possible combinations of $(j_1,k_1;j_2,k_2)$ in the coupling of a
 $(1,1)$ spinless adjoint to a $(1,1)$ vector mode and a zero-mode 
gauge boson, are: $(1,0;1,0)$,
$(2,0;1,1)$, $(1,1;0,0)$ and $(1,1;2,0)$. After summing over such 
contributions, we find,

\begin{equation}
{\cal M} =\frac{1}{4 \pi^{2}}\epsilon^{\mu \nu \alpha \beta}
\epsilon^*_{\nu}(p_2)\epsilon^*_{\mu}(p_3)p_{1\alpha}p_{2\beta}\times
\sum_{\psi}(Gauge Couplings)\; \sigma_{\psi} A_{\psi}
\label{aah_amplit}
\end{equation}

The resulting dimension-5 operators, involving a $(1,1)$ mode spinless
adjoint and two zero-mode gauge bosons, are given in
Eqs.[\ref{su3},\ref{su2}]. The $C$ coefficients arise in
Eqs.[\ref{su3},\ref{su2}] are given by,

\begin{eqnarray}
\textit{C}_{\gamma g}^{G}&=&\frac{1}{4 \pi^2} g_s^2 e~\sum_{\psi_{\pm}} Q_{\psi}\sigma_{\psi}A_{\psi}^{G\gamma g}  \nonumber \\
\textit{C}_{Z g}^{G}&=&\frac{1}{4 \pi^2} g_s^2 \frac{g}{cos \theta_{w}}\sum_{\psi_{\pm}} [I^{3}_{\psi}-Q_{\psi} sin^{2}\theta_{w}]\sigma_{\psi}A_{\psi}^{G Z g} \nonumber \\
\textit{C}_{g g}^{B}&=&\frac{1}{4 \pi^2} g_s^2 g^{\prime}\sum_{\psi_{\pm}} \frac{Y_{\psi}}{2}\sigma_{\psi}A_{\psi}^{B g g}  \nonumber \\
\textit{C}_{\gamma \gamma}^{B}&=&\frac{1}{4 \pi^2} e^2 g^{\prime}\sum_{\psi_{\pm}} \frac{Y_\psi}{2}Q^{2}_{\psi} \sigma_{\psi}A_{\psi}^{B \gamma \gamma}   \nonumber \\
\textit{C}_{\gamma Z}^{B}&=&\frac{1}{4 \pi^2} e g^{\prime}\frac{g}{cos \theta_{w}}\sum_{\psi_{\pm}} \frac{Y_\psi}{2}[I^{3}_{\psi}-Q_{\psi} sin^{2}\theta_{w}] Q_{\psi}\sigma_{\psi}A_{\psi}^{B \gamma Z}  \nonumber \\
\textit{C}_{ZZ}^{B}&=&\frac{1}{4 \pi^2} g^{\prime}\frac{g^2}{cos^{2} \theta_{w}}\sum_{\psi_{\pm}} \frac{Y_\psi}{2}[I^{3}_{\psi}-Q_{\psi} sin^{2}\theta_{w}]^{2} \sigma_{\psi}A_{\psi}^{BZZ} \nonumber \\
\textit{C}_{W^{+}W^{-}}^{B}&=&\frac{1}{4 \pi^2} g^{\prime}\frac{g^2}{2}\sum_{\psi_{\+}} \frac{Y_\psi}{2}\sigma_{\psi}A_{\psi}^{BW^+W^-}  \nonumber \\
\textit{C}_{g g}^{W^{3}}&=&\frac{1}{4 \pi^2} g_s^2 g \sum_{\psi_{+}} I^{3}_{\psi}\sigma_{\psi}A_{\psi}^{W^{3} g g}  \nonumber \\
\textit{C}_{\gamma \gamma}^{W^{3}}&=&\frac{1}{4 \pi^2} e^2 g\sum_{\psi_{+}} I^{3}_{\psi}Q^{2}_{\psi} \sigma_{\psi}A_{\psi}^{W^{3} \gamma \gamma}   \nonumber \\
\textit{C}_{\gamma Z}^{W^{3}}&=&\frac{1}{4 \pi^2} e \frac{g^2}{cos \theta_{w}}\sum_{\psi_{+}} I^{3}_{\psi}[I^{3}_{\psi}-Q_{\psi} sin^{2}\theta_{w}]Q_{\psi} \sigma_{\psi}A_{\psi}^{W^{3} \gamma Z}  \nonumber \\
\textit{C}_{ZZ}^{W^{3}}&=&\frac{1}{4 \pi^2} \frac{g^3}{cos^{2} \theta_{w}}\sum_{\psi_{+}} I^{3}_{\psi}[I^{3}_{\psi}-Q_{\psi} sin^{2}\theta_{w}]^{2} \sigma_{\psi}A_{\psi}^{W^{3} ZZ} \nonumber \\
\textit{C}_{W^{+}W^{-}}^{W^{3}}&=&\frac{1}{4 \pi^2}
\frac{g^3}{2}\sum_{\psi_{+}}
I^{3}_{\psi}\sigma_{\psi}A_{\psi}^{W^{3}W^+W^-} 
\label{C_coeffi}
\end{eqnarray}
Where $\sigma_{\psi_{\pm}}=\pm 1$, $Q_{\psi}$, $I^{3}_{\psi}$ and $Y_{\psi}$ 
are the electric charge, $3^{rd}$ component of isospin and hypercharge
of the corresponding fermion $\psi$ respectively.

It is important to notice that the effective
$A_{H}^{(J_1,K_1)}A_{\nu}^{(J_2,K_2)}A_{\mu}^{(0,0)}$ vertex is
proportional to the gauge anomaly. So in the limit that all the fermions at
each KK level are degenerate in mass, $A_{\psi}$ becomes independent
of $\psi$ and all $C$ coefficients vanish identically due to exact
anomaly cancellation. The mass splittings of KK fermions due to the
radiative corrections thus play a very crucial role for the 
non-zero values of the $C$ coefficients. As for example,
$\textit{C}_{W^{+}W^{-}}^{B}$ is stronger than all other
$\textit{C}^{B}$'s. In case of
$\textit{C}_{W^{+}W^{-}}^{B}$, anomaly cancellation takes place exactly
between $(1,0)$ mode of 6D $+ve$ chirality quarks and lepton
generations. Where as, for others cases, cancellation take place
partially between the $(1,0)$ mode of 6D $+ve$ and $-ve$ chirality quarks
and leptons. The mass splitting between the $(1,0)$ mode $+ve$ 6D
chirality quarks and lepton generations is higher than the mass
splitting between $(1,0)$ mode of 6D $+ve$ and $-ve$ chirality
fermions \cite{dobrescu3}. Similar
kind of argument can be given in favour of the small value of 
$\textit{C}_{W^{+}W^{-}}^{W^{3}}$ compared to all others.

 \end{document}